\documentclass[structabstract]{aa}  
\usepackage{graphicx}
\usepackage{txfonts}
\usepackage{natbib} 
\bibpunct{(}{)}{;}{a}{}{,} 
\usepackage[dvipsnames]{xcolor} 

\newcommand{\um}{\hbox{\,$\mu$m}}
\newcommand{\jybeam}{\hbox{Jy\,beam$^{-1}$}}
\newcommand{\kms}{\hbox{\,km\,s$^{-1}$}}
\newcommand{\ammonia}{\hbox{NH$_3$}}
\newcommand{\msun}{\hbox{\,M$_\odot$}}
\begin{document}
   \title{Two-level hierarchical fragmentation in the Orion Molecular Cloud\,1 northern filament}

   \author{
	P.~S. Teixeira\inst{1}
          \and
          S. Takahashi\inst{2,3} 
          \and
         L.~A. Zapata\inst{4}
          \and
          P.~T.~P. Ho\inst{5}
          }

   \institute{
	Universtit\"at Wien, Institut f\"ur Astrophysik, T\"urkenschanzstrasse 17, 1180 Vienna, Austria, 
              \email{paula.teixeira@univie.ac.at}
         \and
        Joint ALMA Observatory, Alonso de Cordova 3108, Vitacura, Santiago, Chile,
        	     \email{stakahas@alma.cl}
        \and
        National Astronomical Observatory of Japan, 2-21-1 Osawa, Mitaka, Tokyo 181-8588, Japan.
         \and
         Centro de Radioastronom\'{\i}a y Astrof\'{\i}sica, UNAM, Morelia, Mexico,
             \email{lzapata@crya.unam.mx}
        \and
        Academia Sinica Institute of Astronomy and Astrophysics, P.O. Box 23-141, Taipei 10617, Taiwan, 
              \email{pho@asiaa.sinica.edu.tw}
             }

   \date{}


  \abstract
  {The filamentary structure of molecular clouds may set important constraints on the mass distribution of stars forming within them. It is therefore important to understand what physical mechanism dominates filamentary cloud fragmentation and core formation.  }
   {Orion A is the nearest giant molecular cloud, and its ``$\int$- shaped filament'' is a very active star forming region that is a good target for such a study.
      We have recently reported on the collapse and fragmentation properties of the northernmost part of this structure, located $\sim$2.4\,pc north of Orion KL -- the Orion Molecular Cloud\,3 (OMC\,3, Takahashi et al. 2013). As part of our project to study the $\int$-shaped filament, we analyze the fragmentation properties of the northern OMC\,1 filament (located $\lesssim$0.3\,pc north of Orion KL). This filament is a dense structure previously identified by JCMT/SCUBA submillimeter continuum and VLA \ammonia\ observations and shown to have fragmented into clumps. Our aim is to search for cores and young protostars embedded within OMC\,1n, and to study how the filament is fragmenting to form them.}
   {We observed OMC\,1North (hereafter OMC\,1n) with the Submillimeter Array (SMA) at 1.3\,mm and report on our analysis of the continuum data.}
   {We discovered 24 new compact sources, ranging in mass from 0.1 to 2.3, in size from 400 to 1300\,au, and in density from $2.6 \times 10^7$ to $2.8 \times 10^6 $cm$^{-3}$. The masses of these sources are similar to those of the SMA protostars in OMC\,3 (Takahashi et al. 2013), but their typical sizes and densities are lower by a factor of ten. Only 8\% of the new sources have infrared counterparts, yet there are five associated CO molecular outflows. These sources are thus likely in the Class 0 evolutionary phase yet it cannot be excluded that some of the sources might still be pre-stellar cores. 
The spatial analysis of the protostars shows that these are divided into small groups that coincide with previously identified JCMT/SCUBA 850\um\ and VLA NH$_3$ clumps, and that these are separated by a quasi-equidistant length of $\approx$30\arcmin\ (0.06\,pc). This separation is dominated by the Jeans length, and therefore indicates that the main physical process in the filament's evolution was thermal fragmentation. Within the protostellar groups, the typical separation is $\approx$6\arcsec\ ($\sim$2500\,au), which is a factor 2-3 smaller than the Jeans length of the parental clumps within which the protostars are embedded. These results point to a hierarchical (2-level) thermal fragmentation process of the OMC\,1n filament. }
   {}


   \titlerunning{Two-level hierarchical fragmentation in the OMC\,n}
   \authorrunning{Teixeira et al.}
   \maketitle
%
\section{Introduction}
\label{sec:intro}

It is well established that filamentary molecular cloud structures are ubiquitous \citep{schneider79}. 
New observations have generated a revival in the star formation community's interest in these structures, namely observations from the Herschel Space Observatory \citep[e.g.][]{andre10} that revealed the filamentary nature of clouds in great detail. Combined with a better characterization of the spatial structure of clouds, there has also been advancement in the understanding of the kinematics of filamentary structures. Of particular note is the work carried out by \citet{hacar13} where the Taurus filament L1495/B213 was found to have undergone hierarchical fragmentation and is composed of several velocity coherent sub-filaments, each of which has sonic internal velocity dispersion and with a mass-per-unit length consistent with an isothermal cylinder at 10\,K. Hierarchical thermal fragmentation has also been inferred to have occurred in other star forming regions, e.g., in the Orion complex \citep{takahashi13} and in the Spokes cluster in NGC\,2264 \citep{teixeira06,teixeira07,pineda13}. These are much denser and richer star forming environments than Taurus, yet the spatial distribution of protostars within filaments are consistent with the same fragmentation process.

The Orion Molecular Cloud (OMC) A is the nearest \citep[414$\pm$7\,pc,][]{menten07} and one of the richest star-forming giant molecular clouds. It is an elongated structure \citep[``$\int$-shaped filament'',][]{bally87} spanning $\sim$10\,pc along the north-south direction \citep[e.g.][]{johnstone99}, comprising of several large clouds to the north (OMC\,1, OMC\,2, and OMC\,3) and south (OMC\,4 and OMC\,5). These clouds retain the overall filamentary nature of the region and have been the subject of numerous observational studies \citep[see the reviews][and references within]{muench08,o'dell08,peterson08}. 

Of particular interest is the OMC\,1 cloud  for it is the most massive ($>$2200\msun) component of the OMC \citep{bally87} and has spawned the Orion Nebula Cluster \citep[ONC; ][]{o'dell01a}, which is presently located in front of the cloud. 
As shown in Figure \ref{fig:jcmt-sma-fov}, the OMC\,1 cloud is comprised of dense filaments in the north (OMC\,1n) that connect to Orion KL, and of the very active star forming region Orion South \citep[e.g.][]{zapata04a}.
\citet{wiseman98} carried out interferometric NH$_3$ observations of the OMC\,1n filaments  with the VLA using high spectral (0.3\kms) and moderate angular (8\arcsec) resolution; their observations identified filamentary structures throughout a 0.5\,pc region extending to the north from Orion-KL. These northern filaments, henceforth referred to as \hbox{OMC\,1n}, were found to be fragmented into bead-like chains of dense clumps. We re-observed these filaments with the Submillimeter Array (SMA) in order to further characterize their fragmentation and the dense NH$_3$ clumps previously identified. We found 24 previously unknown compact millimeter sources, and their spatially distribution indicates fragmentation of the filament at two length scales. This paper is part of a series of high-angular resolution OMC studies focused on the filamentary structures and their embedded sources, the first of which is \citet{takahashi13} where we discuss the global fragmentation properties of the OMC and of the OMC\,3 cloud in more detail.

We describe our SMA observations of OMC\,1n in \S\,\ref{sec:obs} and present our results in \S\,\ref{sec:res}. Finally, we discuss our findings on the characterization of these new millimeter sources in \S\,\ref{sec:dis} and summarize our conclusions in \S\,\ref{sec:sum}

%
\section{Observations and Data Reduction}
\label{sec:obs}
		
\begin{figure}[!h]
\centering
\includegraphics[width=\columnwidth]{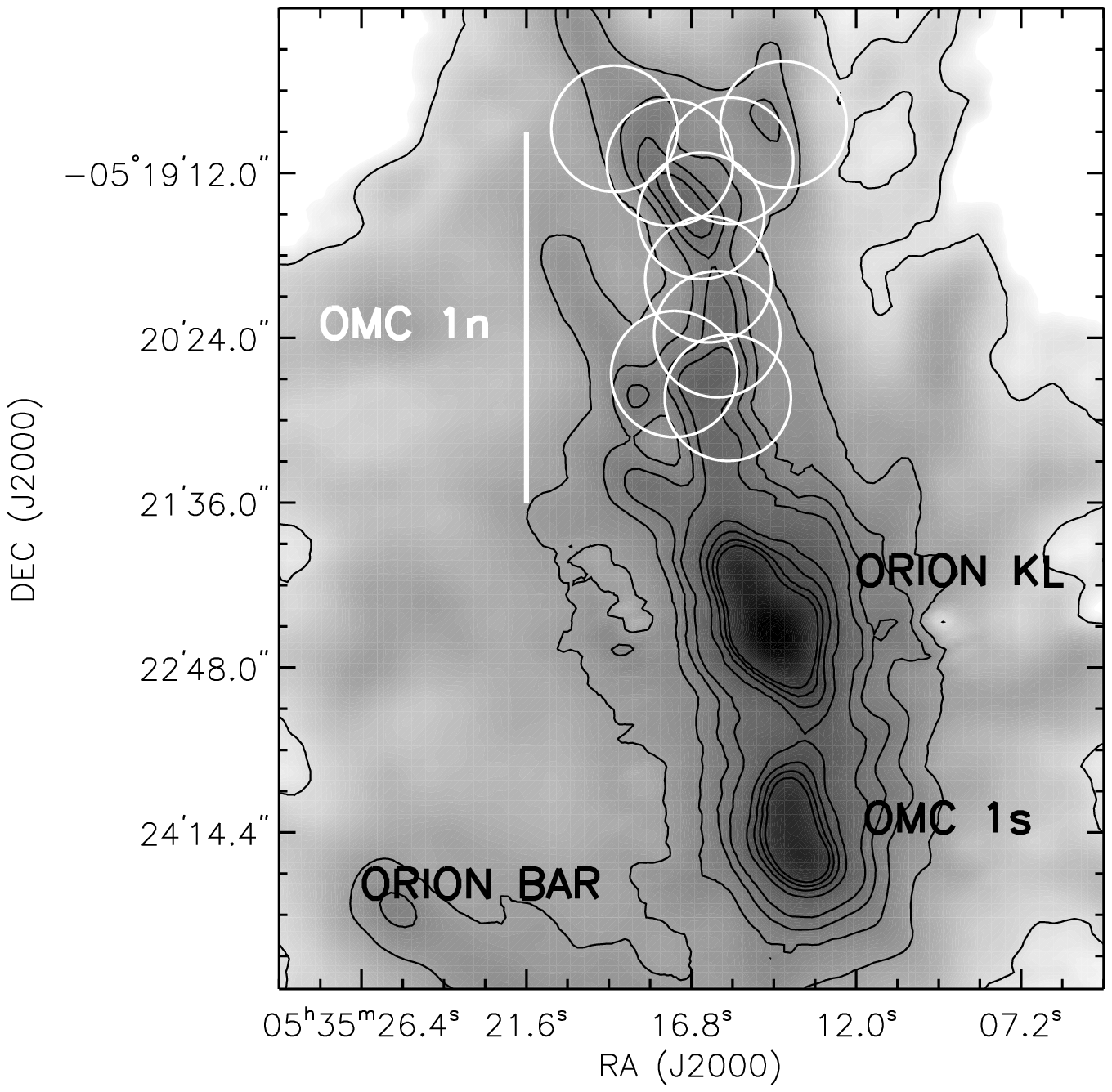}
\caption{JCMT/SCUBA 850\um\ continuum image of the OMC-1 cloud \citep{johnstone99}, with various features indicated as references, such as the Orion Bar, Orion South and Orion\,KL. The black contours range from 0.5 to 8.5\,\jybeam\ in steps of 2\,\jybeam, and from 15 to 30\,\jybeam\ in steps of 5\,\jybeam. The white circles represent the pointings made to cover the SMA area surveyed and their size correspond to the SMA primary beam, 55\arcsec. }
\label{fig:jcmt-sma-fov}
\end{figure}

We observed OMC\,1n using the eight 6\,m antennas of the SMA\footnote{The SMA is a joint project between the SAO and the Academia Sinica Institute of Astronomy and Astrophysics and is funded by the Smithsonian Institution and the Academia Sinica.} \citep{ho04} in the compact configuration on December 4, 20, and 24, 2008 (2007B-S028, P.I.: L.~Zapata) and in the subcompact configuration on February 1, 2009 (2008B-S072, P.I.: L.~Zapata) at 230\,GHz. The primary beam at this frequency is \hbox{55$''$}, and the nine pointing centers were distributed in a Nyquist sampled grid. Figure \ref{fig:jcmt-sma-fov} shows the area surveyed by the SMA where the white circles represent the primary beam size pointings or fields. We observed each field for 1 minute during each of \hbox{30} cycles through all the fields, giving a total on-source integration time of \hbox{30 minutes}. Both the LSB \hbox{(${\nu}_c$=230 GHz)} and USB \hbox{(${\nu}_c$=220 GHz)} data were obtained simultaneously with the \hbox{90$^{\circ}$} phase switching technique by the digital spectral correlator, which had a bandwidth of \hbox{2 GHz} for each sideband. 
		
The SMA correlator covers \hbox{2 GHz} bandwidth in each of the two side-bands separated by \hbox{10 GHz}. Each band is divided into \hbox{24 chunks} of \hbox{104 MHz} width, which were covered by fine spectral resolution (\hbox{256 channels} correspond to a velocity resolution of 0.53\kms\ in our observing setting). In addition to continuum observations, molecular lines such as \hbox{CO (2--1)}, \hbox{$^{13}$CO (2--1),} \hbox{C$^{18}$O(2--1),} and \hbox{$^{13}$CS (5--4),} were simultaneously obtained with the \hbox{2 GHz} bandwidth in each sideband. These molecular line data will be presented in a future article.
	
The continuum data from both sidebands (USB + LSB) were combined to obtain a higher \hbox{signal-to-noise} ratio. The combined configurations of the arrays provided projected baselines ranging from \hbox{5.1 to 56.0\,k${\lambda}$}. Our observations were insensitive to structures larger than 32\arcsec (13000 AU/0.07 pc) at the 10\% level \citep{wilner94}. Passband calibration was achieved by observing the quasars, \hbox{3C 454.3} and \hbox{3C 273}. Amplitudes and phases were calibrated by observations of \hbox{0530+135} and \hbox{0541-056} for the compact configurations and \hbox{0607-085} and \hbox{0541-056} for the subcompact configurations. Those flux densities were determined relative to Uranus. The overall flux uncertainty was estimated to be \hbox{$\sim$20\%}. The pointing accuracy of the SMA observations was \hbox{${\sim}3''$}. The raw data were calibrated using MIR, originally developed for the Owens Valley Radio Observatory \citep{scoville93}  and adopted for the SMA. 
		
In order to produce the mosaic continuum image, certain frequencies (which include aforementioned molecular lines) are subtracted from the visibility data using a miriad task ``uvlin''. The effective bandwidth for the continuum emission is approximately 3.6 GHz. 
We used the task CLEAN in miriad with the ROBUST parameter equal to 2, and then we combined the resulting images
linearly with the task LINMOS. The final synthezised beam size results directly of combining all observations, and is ${\approx}3''$ FWHM (corresponding to 1200 AU). The achieved mean rms noise level for the entire mosaic image is estimated to be \hbox{$\approx$11 mJy}. The SMA observational log for each date and the observational parameters are summarized in Table 1 and Table 2.

\begin{table}[!h]
\centering
\caption{SMA Observational Logs\label{tab:obs-logs}}
\begin{tabular}{lccc}
\hline\hline
Date & Configuration & Number of Antennae & ${\tau}_{225}$ \\
\hline
2007 Dec. 4	& compact 		&  7    &   0.05--0.08 \\
2007 Dec. 20	& compact   	 	&  5	&   0.22--0.33  \\
2007 Dec. 24	& compact          	&  7	&   0.12--0.24   \\
2009 Feb. 1	& subcompact            &  8	&   0.25--0.32   \\
\hline
\end{tabular}
\end{table}

\begin{table}[!h]
\caption{SMA Observational Parameters\label{tab:obs-param}}
\centering
\begin{tabular}{lc}
\hline\hline
Parameter & Value \\
\hline
Configurations                                                  &  3 $\times$ compact and  \\
                                                                & 1 $\times$ subcompact \\
Primary beam HPBW (arcsec)                                      &  55 \\
Grid spacing (arcsec)                                           & 25  \\
Synthesized Beam HPBW (arcsec)                                  &  3.0  \\ 
Equivalent frequency (GHz)                                      & 225 \\
Total continuum bandwidth (GHz)                                 &  3.6 \\ 
Projected base line range (k$\lambda$)                          &  5.1 -- 56.0 \\
Maximum detectable structure (arcsec)\tablefootmark{a}          &  32 \\
Gain calibrators                                                &  0530+135, 0541-056 \\
                                                                &  and 0607-085   \\
Bandpass calibrator                                             &  3C 454.3 and 3C 273 \\
Flux calibrators                                                &  Uranus  \\  
RMS noise level (m\jybeam)                                      &  $\sim$11 \\
\hline
\end{tabular}
\tablefoot{
\tablefoottext{a}{Our observations were insensitive to more extended emission than this size-scale structure at the 10\% level \citep{wilner94}.}
}
\end{table}
		
%
\section{Results}
\label{sec:res}

\subsection{New compact continuum sources}

\begin{figure*}[!ht]
\sidecaption
\includegraphics[width=12cm]{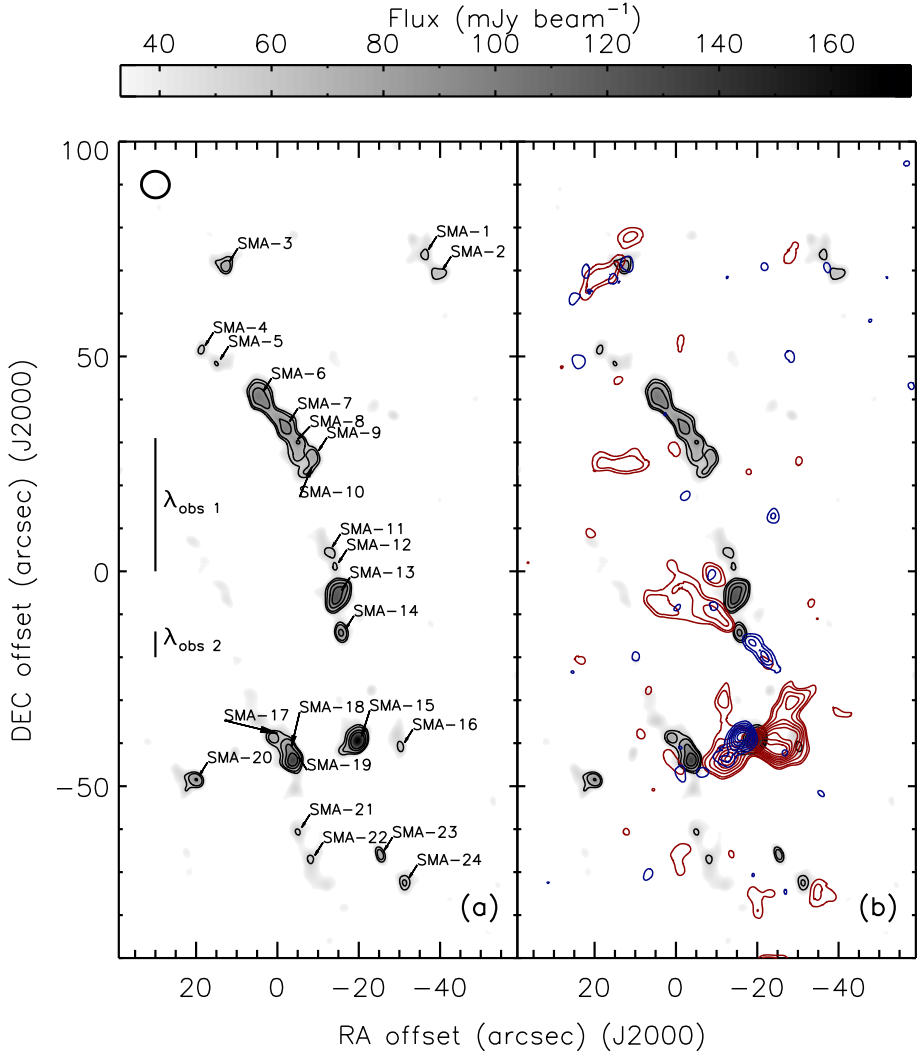}
\caption{SMA 1.3\,mm continuum emission (grayscale, lowest level corresponds to 3-$\sigma$ emission), where the black contours range from 5\,$\sigma$ to 10\,$\sigma$ in steps of 1\,$\sigma$ ($\sigma$ $\sim$ 11\,m\jybeam). Coordinate offsets are measured with respect to ($\alpha, \delta$) = (05$^h$35$^m$17.0$^s$, -05\degr22\arcmin00.0\arcsec) (J2000.0).  In panel (a), the new sources are indicated by arrows with their corresponding IDs (compare with Tables \ref{tab:prop-meas} and \ref{tab:prop-calc}). The vertical scale bars correspond to the median 1st and 7th nearest neighbor separations of the SMA sources, $\lambda_\mathrm{obs\ 1}$ and $\lambda_\mathrm{obs\ 2}$, respectively (see Section \ref{subsubsec:spatial} and Table \ref{tab:lengths}). Panel (b) shows CO (2-1) red- and blue-shifted emission from molecular outflows, in red and blue contours, respectively. The red contour levels range from 14 to 136\,Jy\,beam$^{-1}$\,km\,s$^{-1}$, in steps of 14\,Jy\,beam$^{-1}$\,km\,s$^{-1}$. The blue contour levels range from 9 to 31\,Jy\,beam$^{-1}$\,km\,s$^{-1}$, in steps of 3\,Jy\,beam$^{-1}$\,km\,s$^{-1}$. The molecular outflow velocities range from 0 to 40 (red-shifted emission), and from -20 to 0 (blue-shifted emission).}
\label{fig:sma}
\end{figure*}

We detected 24 previously unknown compact millimeter sources within the OMC-1N region; these new sources were identified as having peak fluxes $\ge$\,5$\sigma$ (i.e. $\ge$\,55m\jybeam) and are shown in Figure \ref{fig:sma}. There is fainter extended emission at the 3$\sigma$ level that may correspond to smaller cores.
Only two of these sources were found to have (nir- and mid) infrared  counterparts; one of them coincides with SMA-10, and the other is located very near SMA-15, more precisely, with the small protrusion extending from the peak of the emission. We also detected red- and/or blue-shifted CO emission arising from bipolar outflows.

We used MIRIAD \citep{sault95} to measure the fluxes, sizes and positions of each source (IMFIT procedure), where each source was modeled by an elliptical 2D Gaussian. As can be seen in Figure \ref{fig:sma}, the sources appear to be distributed in groups, i.e. they are connected as a single structure when using a 3-$\sigma$ contour. We therefore carried out multi-gaussian fits so that all sources in a particular group were fit together. There are six groups of sources that were fit simultaneously: group 1 consists of sources SMA-1 and SMA-2, group 2 consists of sources SMA-4 and SMA-5, group 3 consists of sources SMA-6, SMA-7, SMA-8, SMA-9 and SMA-10, group 4 consists of sources SMA-11, SMA-12, SMA-13 and SMA-14, group 5 consists of sources SMA-17, SMA-18 and SMA-19, and group 6 consists of sources SMA-21 and SMA-22. Figure \ref{fig:sma} shows that SMA-15 may in fact consist of two sources (the presence of an infrared counterpart offset from the position of the main peak and coincident with the small south-west extended emission supports this), however we were unable to fit SMA-15 with two simultaneous 2D gaussians. Likewise for source SMA-20. 
The measured properties for each source are listed in Table \ref{tab:prop-meas}. The total integrated flux listed was measured from the multi-2D-gaussian fitting. Sources SMA-9, SMA-11, and SMA-18 are not well fit by a 2D gaussian model, however it was necessary to include them in order to fit well the nearby brighter compact source(s) within the same group. The errors cited for the total flux correspond to the statistical uncertainties, however they are in fact dominated by the calibration uncertainty that is $\sim$\,20\%, as previously mentioned in \S\,\ref{sec:obs}. The total flux listed in Table \ref{tab:prop-meas} corresponds to thermal dust emission but may also have some contribution from thermal free-free emission from protostellar jets. In order to calculate this latter contribution, we use a spectral index of ~0.6 \citep{panagia75,anglada98} and estimate the flux from thermal free-free emission, $S_\nu$, for these sources from $S_\nu \propto \nu^{0.6}$, where $\nu$ is the frequency of the emission. The SMA sources detected in OMC\,1n share very similar properties with those in OMC\,3 \citep{takahashi13} and OMC\,2 \citep{takahashi15}. We can therefore make use of VLA measurements at 3.6\,cm of OMC\,2 protostars \citep[][ Table 1]{reipurth99} as a benchmark, since most of the flux at this wavelength is from free-free emission. Using their minimum and maximum flux densities, we find that the contribution of thermal free-free emission from jets would range between 1.1\,mJy and 20.9\,mJy at 1.3\,mm for the OMC\,2 sources. We assume that the sources in OMC\,1n would have a similar contribution, however, future observations are necessary to measure the exact amount of  free-free emission for each of the OMC\,1n sources. We note that the estimated flux from thermal free-free emission is typically within the error of the total integrated flux for each source.

\begin{table*}[!ht]
\caption{Measured observables of the SMA sources from 2D elliptical Guassian fitting.}
\label{tab:prop-meas}
\centering
\begin{tabular}{llccccccc}
\hline \hline
 ID 				             & R.A. 		    &  Dec.                 & Pos. unc. & Size    							    & P.A.                    & F$_{230\,GHz\ \mathrm{peak}}$ & F$_{230\,GHz\ \mathrm{int.}}$\tablefootmark{a}  \\
      					    & (J2000.0)       & (J2000.0)         & (arcsec)   & (arcsec)                        			              & (deg.)                & (m\jybeam)                                      & (mJy) \\
\hline
 SMA-1\tablefootmark{c}       & 05:35:14.59 & $-$05:18:46.1 &  0.3 & (3.5\,$\pm$\,1.0)\,$\times$\,(2.9\,$\pm$\,0.9) & $-21\,\pm$\,68 & 60\,$\pm$\,3 & 118\,$\pm$\,50\\
  SMA-2\tablefootmark{c}      & 05:35:14.35 & $-$05:18:50.8 &  0.3 &  (6.9\,$\pm$\,1.7)\,$\times$\,(3.5\,$\pm$\,0.8) & $-79\,\pm$\,17 & 64\,$\pm$\,2 & 224\,$\pm$\,74\\
  SMA-3   				    & 05:35:17.85 & $-$05:18:48.9 &  0.2 &  (4.9\,$\pm$\,1.3)\,$\times$\,(4.3\,$\pm$\,1.2) & $-33\,\pm$\,102 & 76\,$\pm$\,5 & 227\,$\pm$\,88\\
  SMA-4\tablefootmark{d}      & 05:35:18.25 & $-$05:19:08.4 &  0.3 &  (4.5\,$\pm$\,1.7)\,$\times$\,(1.6\,$\pm$\,0.4) & $-23\,\pm$\,23 & 60\,$\pm$\,4 & 112\,$\pm$\,52\\
  SMA-5\tablefootmark{d}      & 05:35:18.00 & $-$05:19:11.7 &  0.3 &  (2.7\,$\pm$\,1.5)\,$\times$\,(1.6\,$\pm$\,0.9) & $22\,\pm$\,100 & 58\,$\pm$\,4 & 83\,$\pm$\,66\\
  SMA-6\tablefootmark{e}      & 05:35:17.29 & $-$05:19:19.1 &  0.2 &  (7.5\,$\pm$\,1.0)\,$\times$\,(5.2\,$\pm$\,0.4) & $23\,\pm$\,12 & 104\,$\pm$\,6 & 485\,$\pm$\,77\\
  SMA-7\tablefootmark{e}      & 05:35:16.87 & $-$05:19:26.1 &  0.2 &  (7.1\,$\pm$\,2.3)\,$\times$\,(4.6\,$\pm$\,0.4) & $36\,\pm$\,13 & 95\,$\pm$\,7 & 389\,$\pm$\,137\\
  SMA-8\tablefootmark{e}      & 05:35:16.67 & $-$05:19:32.3 &  0.3 &  (7.8\,$\pm$\,3.4)\,$\times$\,(4.4\,$\pm$\,1.1) & $-51\,\pm$\,28 & 66\,$\pm$\,22 & 290\,$\pm$\,174\\
  SMA-9\tablefootmark{b,e}   & 05:35:16.39 & $-$05:19:32.9 &  0.4 &  (3.2\,$\pm$\,1.8)\,$\times$\,(2.6\,$\pm$\,1.6)  & $4\,\pm$\,71 & 39\,$\pm$\,25 & 30\,$\pm$\,31\\
  SMA-10\tablefootmark{e}    & 05:35:16.50 & $-$05:19:36.7 &  0.3 &  (8.0\,$\pm$\,3.4)\,$\times$\,(2.4\,$\pm$\,0.9) & $-53\,\pm$\,16 & 59\,$\pm$\,19 & 197\,$\pm$\,129\\
  SMA-11\tablefootmark{f}     & 05:35:16.17 & $-$05:19:55.3 &  0.3 &  (5.6\,$\pm$\,1.5)\,$\times$\,(4.4\,$\pm$\,1.4) & $74\,\pm$\,133 & 61\,$\pm$\,7 & 204\,$\pm$\,86\\ 
  SMA-12\tablefootmark{b,f}  & 05:35:16.06 & $-$05:19:59.0 &  0.5 &  (3.1\,$\pm$\,2.1)\,$\times$\,(2.4\,$\pm$\,1.8) & $-29\,\pm$\,70 & 32\,$\pm$\,27 & 22\,$\pm$\,30\\
  SMA-13\tablefootmark{f}     & 05:35:16.01 & $-$05:20:05.4 &  0.1 &  (7.0\,$\pm$\,0.4)\,$\times$\,(4.4\,$\pm$\,0.2) & $-30\,\pm$\,6 & 128\,$\pm$\,5 & 509\,$\pm$\,44\\ 
  SMA-14\tablefootmark{f}     & 05:35:15.95 & $-$05:20:14.4 &  0.2 &  (3.6\,$\pm$\,0.4)\,$\times$\,(2.6\,$\pm$\,0.3) & $5\,\pm$\,20 & 91\,$\pm$\,7 & 171\,$\pm$\,33\\
  SMA-15                                  & 05:35:15.69 & $-$05:20:39.4 &  0.1 &  (3.3\,$\pm$\,0.1)\,$\times$\,(2.3\,$\pm$\,0.1) & $-32.\,\pm$\,6 & 173\,$\pm$\,2 & 301\,$\pm$\,14\\
  SMA-16                                  & 05:35:14.98 & $-$05:20:40.8 &  0.3 &  (5.5\,$\pm$\,1.6)\,$\times$\,(2.4\,$\pm$\,0.6) & $14\,\pm$\,17 & 61\,$\pm$\,1 & 146\,$\pm$\,55\\
  SMA-17\tablefootmark{g}   & 05:35:17.07 & $-$05:20:38.7 &  0.2 &  (3.6\,$\pm$\,0.9)\,$\times$\,(1.9\,$\pm$\,0.3) & $-13\,\pm$\,16 & 74\,$\pm$\,6 & 213\,$\pm$\,66 \\
  SMA-18\tablefootmark{g}   & 05:35:16.84 & $-$05:20:41.8 &  0.3 &  (10.3\,$\pm$\,13.7)\,$\times$\,(1.1\,$\pm$\,0.6) & $-20\,\pm$\,18 & 55\,$\pm$\,70 & 186\,$\pm$\,358\\
 SMA-19\tablefootmark{g}    & 05:35:16.72 & $-$05:20:44.2 &  0.1 & (5.2\,$\pm$\,1.9)\,$\times$\,(3.8\,$\pm$\,2.7) & $63\,\pm$\,64 & 118\,$\pm$\,37 & 199\,$\pm$\,171 \\
 SMA-20                                   & 05:35:18.35 & $-$05:20:48.5 &  0.2 &  (4.7\,$\pm$\,0.8)\,$\times$\,(2.9\,$\pm$\,0.5) & $75\,\pm$\,25 & 86\,$\pm$\,5 & 207\,$\pm$\,49\\
  SMA-21\tablefootmark{h}   & 05:35:16.67 & $-$05:21:00.6 &  0.3 &  (5.8\,$\pm$\,2.6)\,$\times$\,(3.5\,$\pm$\,1.6) & $-2\,\pm$\,56 & 57\,$\pm$\,1 & 170\,$\pm$\,107\\
  SMA-22\tablefootmark{h}   & 05:35:16.45 & $-$05:21:07.1 &  0.3 &  (4.1\,$\pm$\,1.1)\,$\times$\,(3.7\,$\pm$\,1.3) & $-13\,\pm$\,99 & 58\,$\pm$\,2 & 141\,$\pm$\,63\\
  SMA-23\tablefootmark{b}   & 05:35:15.31 & $-$05:21:05.9 &  0.2 &  (5.3\,$\pm$\,0.6)\,$\times$\,(3.2\,$\pm$\,0.3) & $24\,\pm$\,7 & 76\,$\pm$\,4 & 120\,$\pm$\,19\\
  SMA-24                                 & 05:35:14.91 & $-$05:21:12.3 &  0.3 &  (5.3\,$\pm$\,0.7)\,$\times$\,(4.2\,$\pm$\,0.5) & $-4\,\pm$\,21 & 66\,$\pm$\,3 & 203\,$\pm$\,36\\
\hline
\end{tabular}
\tablefoot{Units of right ascension are hours, minutes, and seconds, and units of declination are degrees, arcminutes, and arcseconds.\\
\tablefoottext{a}{Total integrated flux. We estimate the thermal free-free contribution to range between 1.1\,mJy and 20.9\,mJy (see text for details). This value is typically smaller than the error of F$_{230\,GHz\ \mathrm{int.}}$ }\\
\tablefoottext{b}{un-deconvolved.}\\
\tablefoottext{c}{simultaneous fit of the group of sources SMA-1 and SMA-2.}\\
\tablefoottext{d}{simultaneous fit of the group of sources SMA-4 and SMA-5.}\\
\tablefoottext{e}{simultaneous fit of the group of sources SMA-6, SMA-7, SMA-8, SMA-9, SMA-10.}\\
\tablefoottext{f}{simultaneous fit of the group of sourcesSMA-11, SMA-12, SMA-13, and SMA-14.}\\
\tablefoottext{g}{simultaneous fit of the group of sources SMA-17, SMA-18, and SMA-19.}\\
\tablefoottext{h}{simultaneous fit of the group of sources SMA-21 and SMA-22.}
}
\end{table*}

Before proceeding, we would like to make a brief remark on the nomenclature used in this paper. The terms "core" and "clump" do not have a strict definition in the field of star formation and are sometimes used interchangeably in the literature, which may lead to erroneous comparisons of results. We use these two terms in the following manner: a core (or several cores) is formed within a clump and is therefore relatively denser than the clump, and we follow the nomenclature used in \citet{takahashi13}, where "small-scale clumps" in the $\int$-shaped-filament are structures with sizes on the order of 0.3-0.1\,pc, and "cores" are structures with sizes typically 10 times smaller. The aforementioned groups of sources would therefore correspond to small-scale clumps that fragmented into cores, e.g. group 1 corresponds to a single small-scale clump that fragmented into two cores, group 3 corresponds to a single small-scale clump that fragmented into five cores, etc. Each core corresponds to an SMA source.

\begin{figure*}[!ht]
\sidecaption
\includegraphics[width=12cm]{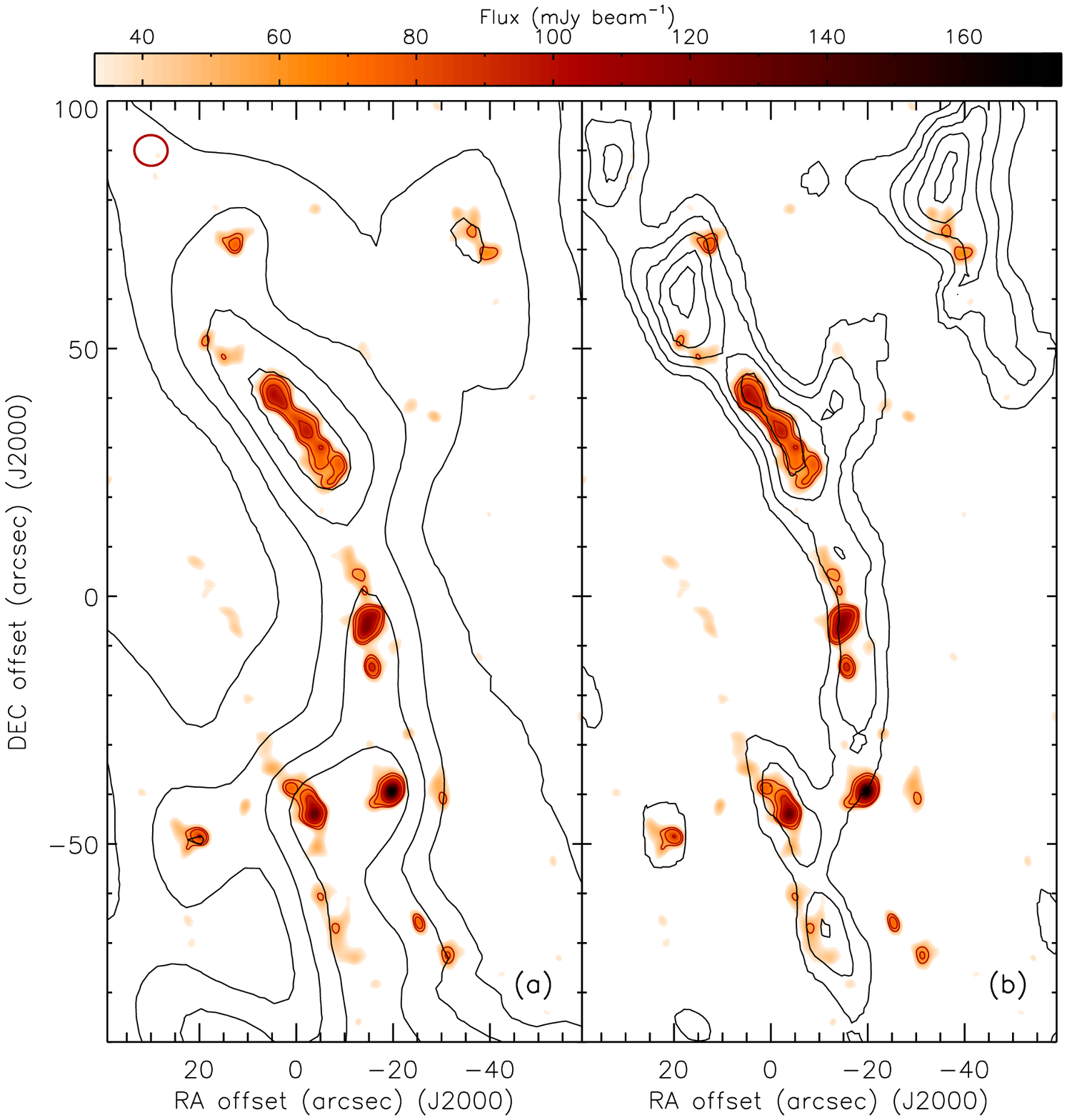}
\caption{Comparison of SMA continuum emission (grayscale) with: (a) 850\um\ JCMT emission, where the JCMT contours range from 0.5 to 20\,\jybeam\ in steps of 1\,\jybeam\ \citep{johnstone99}, and (b) VLA NH$_3$ (1,1) emission integrated from 4.6 to 12.3\kms, where two lowest NH$_3$ contours are 20 and 100\,m\jybeam\kms\ and the higher level contours continue in steps of 100\,m\jybeam\kms\ \citep{wiseman98}. Coordinate offsets are measured with respect to ($\alpha, \delta$) = (05$^h$35$^m$17.0$^s$, -05\degr22\arcmin00.0\arcsec) (J2000.0). \emph{[This figure is available in color in the online electronic journal.]}  }
\label{fig:sma-jcmt-vla}
\end{figure*}

The left panel of Figure \ref{fig:sma-jcmt-vla} compares the SMA 1.3\,mm sources (grayscale) with previously observed JCMT/SCUBA 850\um\ emission \citep{johnstone99} (contours).  Our SMA observations spatially resolved all the JCMT/SCUBA clumps, i.e. all of the single-dish identified clumps are associated with at least one SMA source. The right panel of Figure \ref{fig:sma-jcmt-vla} compares the SMA 1.3\,mm sources (grayscale) with previously observed VLA NH$_3$ (1,1) emission \citep{wiseman98} (contours). The SMA sources clearly lie within dense filamentary regions, although they are not always located at the center of an ammonia clump. For example, sources SMA-1 and SMA-2 are located at the center of the same 850\um\ clump, yet are spatially offset by $\sim$10\,\arcsec\ from the center of an ammonia core. Other sources that are not located near the center of ammonia small-scale clumps are SMA-3, SMA-4, SMA-5, SMA-13, SMA-16, SMA-21, SMA-23, and SMA-24. 

\subsection{Masses and sizes}
\label{subsec:mass}

We can determine the masses of these sources if we assume the 1.3\,mm emission arises from optically thin dust that can be characterized by a single temperature \citep[e.g.][]{mundy95,bally98}. The total gas+dust mass is calculated from the following equation:

\begin{equation}
M = \frac{F_\nu d^2}{B_\nu(T_d) \kappa_\nu},
\label{eq:mass}
\end{equation}

\noindent where $F_\nu$ is the measured total flux of the source, $d$ is the distance to the source, $B_\nu(T_d)$ is the Planck function for a dust temperature $T_d$, and $\kappa_\nu$ is the dust mass opacity. We follow \citet{menten07} and adopt a distance of 414$\pm$7\,pc to the SMA sources. The dust mass opacity is calculated from $\kappa_\nu = 0.1(\nu / 1000\,\mathrm{GHz})^\beta\,\mathrm{cm}^2\,\mathrm{g}^{-1}$ \citep{beckwith90}, assuming a gas-to-dust ratio of 100, and where we take the emissivity index, $\beta$, to be 1.5, following \citet{johnstone99} and the latest results from Planck for this region \citep{lombardi14}; $\kappa_\nu$ is therefore 
$\kappa_{230\,GHz} = 11\times10^{-3}\,\mathrm{cm}^2\,\mathrm{g}^{-1}$. 
According to \citet{wiseman98}, the filament within which the SMA sources are embedded in has a kinetic gas temperature of 15\,K, and this is the temperature we assume for $T_d$ in our mass calculation (assuming that the gas and dust temperatures are coupled). 
Using the integrated flux densities given in Table \ref{tab:prop-meas}, we calculate the masses for each source and their values are shown in Table \ref{tab:prop-calc}. We calculated the radius of each source by taking the geometric mean of the deconvolved major and minor axes (given in Table\,\ref{tab:prop-meas}), \hbox{$R_{geom} = \sqrt{size_{maj.\ axis}\ \cdot\ size_{min.\ axis}}$}, which was in turn used to determine the number densities, $\bar{n}_{H_2}$, for each source assuming spherical symmetry. Both $R_{geom}$ and $\bar{n}_{H_2}$ are shown in Table\,\ref{tab:prop-calc}.

\citet{tatematsu08} find a constant temperature of $\sim$20\,K over the entire $\int$-shaped filament. Here we opted for using the temperature measured by \citet{wiseman98} because this value was obtained from higher angular resolution interferometric observations. The single-dish observations from \citet{tatematsu08} include extended emission of the cloud that envelope the small-scale clumps and that is presumably warmer, thus providing a slightly higher temperature. This effect is also suggested by Herschel data \cite[e.g. see Fig. 2 of ][]{lombardi14}. On the other hand, the interferometric observations filtered out the extended emission and should give a more reliable estimate of the temperature of the dense regions within the filament. For comparison purposes we added a note to Table\,\ref{tab:prop-calc} where we show that changing the temperature assumption to 20\,K would lower the masses by factor of $\sim$1.5.

\begin{table}[!ht]
\caption{Calculated properties of the SMA sources and their nearest neighbor separation.}
\label{tab:prop-calc}
\centering
\begin{tabular}{lccccccc}
\hline \hline
ID & R$_\mathrm{geom}$ & Mass\tablefootmark{a} & $\bar{n}_{H_2}$ & NNS \\
    & (AU) & (M$_\odot$) & (10$^{7}$\,cm$^{-3}$) & (arcsec)\\
\hline
SMA-1  & 661\,$\pm$\,195 & 0.5\,$\pm$\,0.2 & 6.6\,$\pm$\,3.4 & 5.8 $\pm$ 0.6\\
SMA-2  & 1012\,$\pm$\,234 & 1.0\,$\pm$\,0.3 & 3.5\,$\pm$\,1.4 & 5.8 $\pm$ 0.6 \\
SMA-3  & 949\,$\pm$\,258 & 1.0\,$\pm$\,0.4 & 4.3\,$\pm$\,2.0 & 20.5 $\pm$ 0.5 \\
SMA-4  & 557\,$\pm$\,177 & 0.5\,$\pm$\,0.2 & 10.5\,$\pm$\,5.9 & 4.9 $\pm$ 0.6 \\
SMA-5  & 436\,$\pm$\,243 & 0.4\,$\pm$\,0.3 & 16.3\,$\pm$\,15.8 & 4.9 $\pm$ 0.6 \\
SMA-6  & 1285\,$\pm$\,124 & 2.2\,$\pm$\,0.3 & 3.7\,$\pm$\,0.7 & 9.5 $\pm$ 0.4 \\
SMA-7  & 1178\,$\pm$\,210 & 1.7\,$\pm$\,0.6 & 3.9\,$\pm$\,1.5 & 6.9 $\pm$ 0.5 \\
SMA-8  & 1217\,$\pm$\,395 & 1.3\,$\pm$\,0.8 & 2.6\,$\pm$\,1.8 & 4.2 $\pm$ 0.7 \\
SMA-9  & 594\,$\pm$\,343 & 0.1\,$\pm$\,0.1 & 2.3\,$\pm$\,2.8 & 4.2 $\pm$ 0.7 \\
SMA-10 & 906\,$\pm$\,368 & 0.9\,$\pm$\,0.6 & 4.3\,$\pm$\,3.3 & 4.2 $\pm$ 0.7 \\
SMA-11 & 1024\,$\pm$\,292 & 0.9\,$\pm$\,0.4 & 3.1\,$\pm$\,1.6 & 4.0 $\pm$ 0.8 \\
SMA-12 & 564\,$\pm$\,407 & 0.1\,$\pm$\,0.1 & 2.0\,$\pm$\,3.0 & 4.0 $\pm$ 0.8 \\
SMA-13 & 1151\,$\pm$\,63 & 2.3\,$\pm$\,0.2 & 5.4\,$\pm$\,0.6 & 6.5 $\pm$ 0.6 \\
SMA-14 & 629\,$\pm$\,79 & 0.8\,$\pm$\,0.1 & 11.1\,$\pm$\,2.6 & 9.0 $\pm$ 0.3 \\
SMA-15 & 570\,$\pm$\,18 & 1.3\,$\pm$\,0.1 & 26.4\,$\pm$\,1.5 & 10.6 $\pm$ 0.4 \\
SMA-16 & 757\,$\pm$\,199 & 0.7\,$\pm$\,0.2 & 5.5\,$\pm$\,2.5 & 10.6 $\pm$ 0.4 \\
SMA-17 & 534\,$\pm$\,109 & 1.0\,$\pm$\,0.3 & 22.6\,$\pm$\,8.4 & 3.0 $\pm$ 0.5 \\
SMA-18 & 694\,$\pm$\,587 & 0.8\,$\pm$\,1.6 & 9.0\,$\pm$\,18.9 & 3.0 $\pm$ 0.5 \\
SMA-19 & 917\,$\pm$\,470 & 0.9\,$\pm$\,0.8 & 4.2\,$\pm$\,4.2 & 4.7 $\pm$ 0.4 \\
SMA-20 & 771\,$\pm$\,124 & 0.9\,$\pm$\,0.2 & 7.3\,$\pm$\,2.1 & 23.6 $\pm$ 0.5 \\
SMA-21 & 939\,$\pm$\,417 & 0.8\,$\pm$\,0.5 & 3.3\,$\pm$\,2.6 & 7.2 $\pm$ 0.5 \\
SMA-22 & 810\,$\pm$\,249 & 0.6\,$\pm$\,0.3 & 4.3\,$\pm$\,2.3 & 7.2 $\pm$ 0.5 \\
SMA-23 & 850\,$\pm$\,89 & 0.5\,$\pm$\,0.1 & 3.2\,$\pm$\,0.6 & 8.8 $\pm$ 0.5 \\
SMA-24 & 969\,$\pm$\,116 & 0.9\,$\pm$\,0.2 & 3.6\,$\pm$\,0.8 & 8.8 $\pm$ 0.5 \\
\hline
\end{tabular}
\tablefoot{
\tablefoottext{a}{If a warmer  $T_d$ of 20\,K is used, then the masses of the sources are lower by a factor of $\sim$1.5. Taking into account the estimated thermal free-free emission from jets, the masses listed may be lower by a value smaller than 0.1\msun.}\\
}
\end{table}

\subsection{Spatial distribution}
\label{subsubsec:spatial}

\begin{figure*}[!ht]
\sidecaption
\includegraphics[width=12cm]{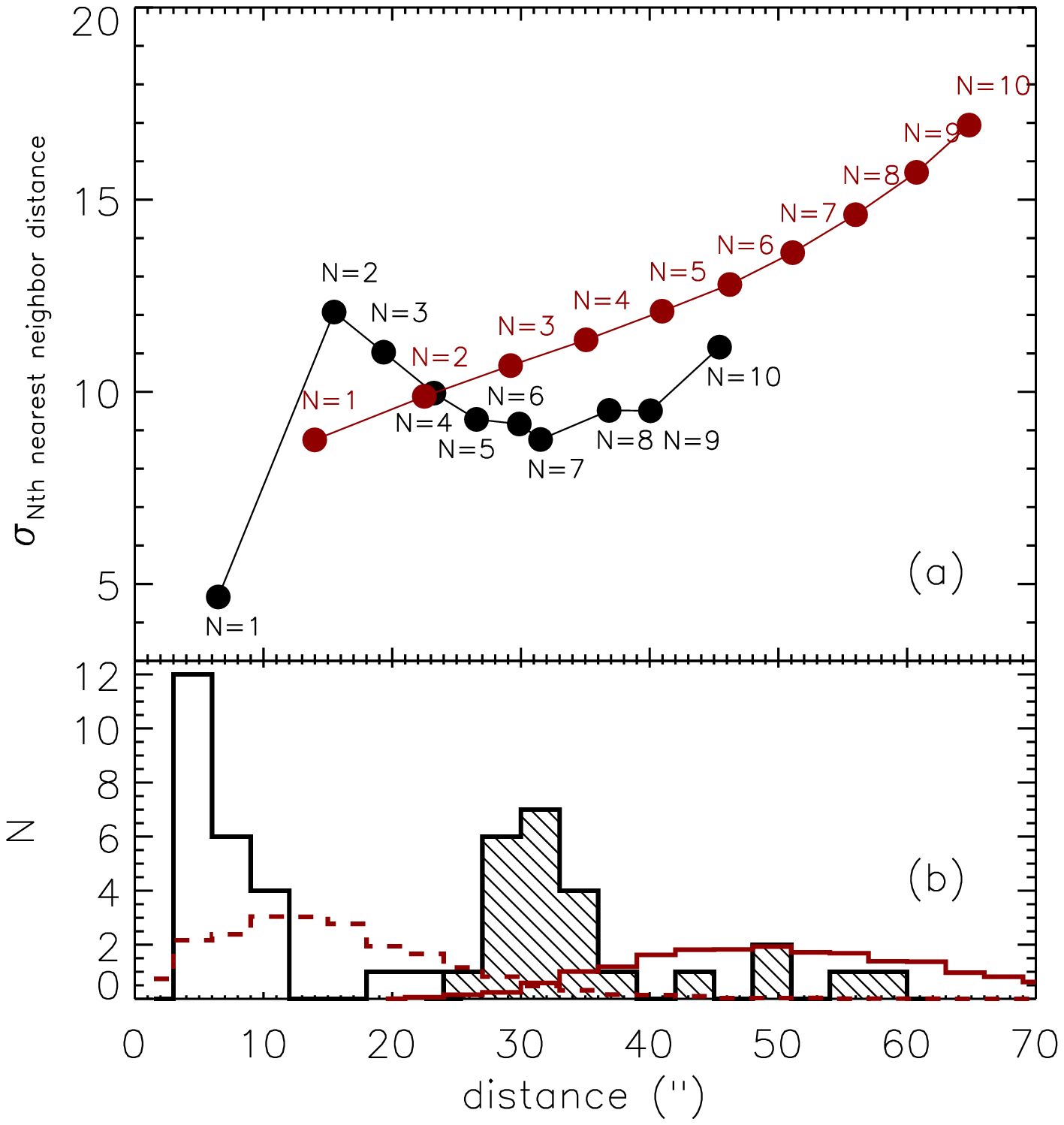}
\caption{(a) Median Nth nearest neighbor separations of the SMA sources vs. $\sigma_\mathrm{Nth\ NNS}$, where each data point (black filled symbol) is labeled from \hbox{N = 1,... 10.} The red symbols represent the median Nth NNS for a random spatial distribution for comparison purposes (see text). 
(b) Distribution of the 1st NNS (open black histogram), and the distribution of the 7th NNS (filled black histogram). The red dashed-line and the red solid-line histograms correspond to the 1st and 7th NNS for random spatial distributions, respectively.
All the histogram bin widths correspond to the approximate SMA synthesized beamsize, 3\arcsec.}
\label{fig:separations}
\end{figure*}

Figure \ref{fig:sma-jcmt-vla} shows that the SMA sources are distributed along dense filamentary structures, and part of groups that show a quasi-equal spacing between (and within) them. To better describe quantitatively these spatial scales associated with the SMA sources, we present in Figure \ref{fig:separations} an analysis of the 1st to the 10th (projected) nearest neighbor separations (NNS). Panel (a) shows the measured median value of the Nth (\hbox{N = 1,... 10}) NNS in black vs. their $\sigma$-value. The plot shows that the empirical function has two minima, namely, for the 1st and 7th nearest neighbors.  A minima in the $\sigma_{Nth\ NNS}$ function indicates a characteristic spatial scale of the distribution of sources (i.e., the width of the NNS distribution narrows as sources have similar separation values). Panel (b) shows that the distributions of the 1st nearest neighbor distribution (open, black line histogram ) peaks at the median value of 6.5\arcsec\  with a $\sigma$ of 4.7\arcsec; the 7th nearest neighbor distribution (filled, black line histogram) peaks at the median value of 31.5\arcsec\ with a $\sigma$ of 8.8\arcsec \footnote{For a sample of sources that excludes SMA-9, -12, and -18, we find that the 1st\ NNS has a median value of 7.6\arcsec\ with a $\sigma$ of 4.3\arcsec, and the 7th NNS has a median value of 38.6\arcsec\ and a corresponding $\sigma$ of 9.4\arcsec. Although the measured parameters of these sources have larger error bars (c.f. Table \ref{tab:prop-meas}), they are fully consistent with the statistical trend of the other sources.}

 In order to determine if these two distributions correspond to bona fide characteristic spatial scalings of the region, we carried out 10,\,000 Monte Carlo simulations of a randomly spatially distributed sample of sources, of the same number as our SMA detected sources, and located within an area of the same size. Panel (a) of Figure \ref{fig:separations} shows how the median separations for this randomly spatial distributed sample (red) increases monotonically, presenting only one minima, for N=1 (we found that the $\sigma_{Nth\ NNS}$ function for a random distribution always increases monotonically, irrespective of the number of sources or the size of the area in which they are distributed). The 1st NNS for the simulated sample has a median value of 14.0\arcsec\ and $\sigma$ of 8.6\arcsec, which is greater than the homologous measured values by a factor larger than 2.  Panel (b) shows how the randomly spatial distributed sample presents broad and shallow peaks for their 1st and 7th nearest neighbor distributions (red) that cannot describe the narrow, well defined, peaks of the observed distributions of the SMA sources. The two spatial scalings we measure are therefore not random and are a characteristic feature of OMC\,1n. We also carried out a two-point correlation function and found positive correlation at the same two spatial scales (see Figure \ref{fig:correlation}). We continue our analysis assuming these spatial scalings are a result of the fragmentation of the parental structures that gave rise to the small-scale clumps and cores in OMC\,1n.

%
\section{Discussion}
\label{sec:dis}

The size and density of the new SMA sources, combined with a general lack of counterpart infrared sources and the presence of CO outflows, 
indicate these compact SMA sources are either cores with young protostars in the Class\,0 evolutionary phase or pre-stellar cores. From \citet{takahashi13} we know the compact submillimeter sources in OMC\,3 range in masses from 0.3\msun\ to 5.7\msun, and we find in this work that the masses of the OMC\,1n sources range between 0.1\msun\ and 3.0\msun. The OMC\,3 projected source sizes range from 1400\,au to 8200\,au, whereas the projected OMC\,1n source sizes range from 400\,au to 1300\,au. In terms of density, sources in OMC\,3 have \hbox{$2.0 \times 10^8$\,cm$^{-3} >\  n_{H_2} > 1.9 \times 10^6$\,cm$^{-3} $}, whereas the OMC\,1n sources are on average less denser, and have \hbox{$2.6 \times 10^7$\,cm$^{-3} > n_{H_2} > 2.8 \times 10^6$\,cm$^{-3} $}. Finally, comparison of the fraction of infrared counterparts, 67\% for OMC\,3 and 8\% for OMC\,1n, indicates that the latter are less evolved than the former.  We find that the OMC\,1n sources are indeed very young objects that have not had enough time to substantially move away from their birth site within the filament. Class\,0 protostars \citep[$\lesssim10^4$\,yr;][]{larson03} with an average velocity of 1\,km\,s$^{-1}$ may travel upto $\sim$0.01\,pc away from their birth site during this evolutionary phase. As such, it is valid to infer the fragmentation scale of the parental filament from the measured spatial distribution of these very young sources.

The measured separations are projected distances. The observed scales, $\lambda_\mathrm{obs}$, would therefore correspond to the fragmentation scales, $\lambda_\mathrm{frag}$, if the inclination, $i$, of the filament and small-scale clumps with respect to the line-of-sight is 90\degr\ [$\lambda_\mathrm{obs}$ = $\lambda_\mathrm{frag} \cdot sin(i)$]. 
As discussed in section \S\,\ref{subsubsec:spatial} and shown in Figure \ref{fig:separations}, there are two observed spatial scales present in OMC\,1n: a larger scale corresponding to $\lambda_\mathrm{obs\_1}$ = 31\arcsec\ $\pm$ 9\arcsec\ (0.06 $\pm$ 0.02\,pc), and a smaller scale corresponding to $\lambda_\mathrm{obs\_2}$ = 6\arcsec\ $\pm$ 5\arcsec\ (0.01 $\pm$ 0.01\,pc). The $\lambda_\mathrm{obs\_1}$ scale corresponds to the distance between the JCMT/SCUBA dust emission small-scale clumps and to the distance between the VLA \ammonia\ small-scale clumps (see Figure \ref{fig:sma-jcmt-vla}). The $\lambda_\mathrm{obs\_2}$ scale corresponds to the separation of the new SMA compact sources within these small-scale clumps.  
The inclination of the OMC\,1n filament is unknown. The median inclination of a randomly oriented filament is 60\degr\ \citep{hanawa93}; using this angle of inclination the fragmentation scales would be $\lambda_\mathrm{frag\_1}$=36\arcsec\  (0.07\,pc) and $\lambda_\mathrm{frag\_2}$=7\arcsec\ (0.01\,pc). The projection effect is thus smaller than the error in $\lambda_\mathrm{obs\_1}$ and $\lambda_\mathrm{obs\_2}$  for an inclination of 60\degr. In fact, the projection effect is only significant if the inclination angle is smaller than 50\degr.

A filament may be supported against gravitational collapse and fragmentation through different physical means, of which we consider the following three: thermal pressure, turbulent pressure, and magnetic pressure. These physical processes may affect differently the final fragmentation scale of the filament. In addition, the kinematics of the filament and small-scale clumps may also imprint on the spatial distribution of the sources formed therein. The filament could be experiencing large-scale collapse. Within the filament, the small-scale s could also be undergoing local collapse. The large scale collapse of the filament may shorten the distance between the small-scale clumps, and likewise, local collapse of the small-scale clumps may shorten the distance between the cores therein. The measured fragmentation scales may thus give us insight into which of these aforementioned processes are dominant. We discuss briefly each of these scenarios. 

Let us first consider at what spatial scales the filament would fragment if no support against gravitational collapse is provided by either turbulence or magnetic fields, i.e., if we were to consider only thermal pressure. The Jeans length \citep{spitzer98}, $\lambda_\mathrm{Jeans}$, given by:

\begin{equation}
\lambda_\mathrm{Jeans} = \sqrt{ \frac{\pi c_s^2}{G \rho_0} }
\label{eq:jeans}
\end{equation}

\noindent where $c_s$ is the sound speed given by $\sqrt{(k_B T)/(\mu m_H)}$ ($k_B$ is the Boltzmann constant, $T$ is the temperature, $\mu$ is the mean molecular weight, and $m_H$ is the proton mass), $G$ is the gravitional constant, and $\rho_0$ is the initial volume density given by $n_0/(\mu m_H)$.
We calculated the Jeans length of the filament, $\lambda_\mathrm{Jeans\_1}$, by
using the peak density of the OMC\,1n filament \citep[$n_{H_2}\sim$10$^5$\,cm$^{-3}$, ][]{johnstone99} and a temperature of  15\,K  \citep{wiseman98}. For these conditions the local Jeans length for the filament is 0.08\,pc (40\arcsec). For a temperature of 10\,K, $\lambda_\mathrm{Jeans\_1}$ = 0.07\,pc (35\arcsec). The larger fragmentation scale measured $\lambda_\mathrm{obs\_1}$ is thus consistent with $\lambda_\mathrm{Jeans\_1}$ for inclinations angles greater than 50\degr. It is therefore possible that the OMC\,1n filament underwent thermal fragmentation giving origin to the small-scale clumps therein. For the analysis of the fragmentation of the small-scale clumps themselves, we calculated the local Jeans length using Equation \ref{eq:jeans}, a density of 5\,$\times$\,10$^5$\,cm$^{-3}$ \citep{wiseman98}, and a temperature of 15\,K as before: $\lambda_\mathrm{Jeans\_2}$ = 0.033\,pc (16\arcsec). For a temperature of 10\,K (canonical value of the temperature of a dense pre-stellar region, i.e., that has not yet undergone internal heating), $\lambda_\mathrm{Jeans\_2}$ = 0.026\,pc (13\arcsec). The measured separation between individual cores within the small-scale clumps $\lambda_\mathrm{obs\_2}$ is smaller than $\lambda_\mathrm{Jeans\_2}$ by a factor of $\sim$ 2-3. The measured spatial lengths and Jeans lengths are summarized in Table \ref{tab:lengths}.
 
 \begin{table}[!h]
 \caption{Summary of measured spatial scales and corresponding local Jeans lengths.}
 \label{tab:lengths}
 \begin{tabular}{lll}
 \hline
 \hline
Parental & Measured Scale & Local Jeans Length\\
Structure &  & \\
 \hline
 filament & $\lambda_\mathrm{obs\_1}$ = 31\arcsec\ $\pm$ 9\arcsec  &    $\lambda_\mathrm{Jeans\_1}$(10\,K) = 35\arcsec (0.07\,pc)\\
 	       &	& 	$\lambda_\mathrm{Jeans\_1}$(15\,K) = 40\arcsec (0.08\,pc)\\
\hline	       
small-scale & $\lambda_\mathrm{obs\_2}$ = 6\arcsec\ $\pm$ 5\arcsec & $\lambda_\mathrm{Jeans\_2}(10\,K)$ = 13\arcsec (0.026\,pc)\\
clump                   &          & $\lambda_\mathrm{Jeans\_2}$(15\,K) = 16\arcsec (0.033\,pc)\\
 \hline
 \end{tabular}
 \end{table}
 
For comparison, our previous SMA study of the OMC\,3 region in the \hbox{$\int$-shaped} filament \citep{takahashi13} detected 12 spatially resolved continuum sources at 850\um. We measured a quasi-periodic separation between these individual sources of $\sim$17\arcsec\ (0.035\,pc). This spatial scale is smaller than the local thermal Jeans length by a factor of 2.0-7.3.

Our finding -- \emph{2 level hierarchical fragmentation in OMC\,1n} -- is in general agreement with studies in other star forming regions, notably in the Spokes cluster in NGC\,2264 \citep{teixeira06, teixeira07}, in the more massive infrared dark cloud IRDC G11.11-0.12 \citep{kainulainen13}, and in other massive star forming regions \citep{palau15}. Further observational evidence for this type of fragmentation is given in a recent study of the young stellar population of the Orion\,A and B clouds, where \citet{megeath15} find that the median separation between sources is similar to the Jeans length. In all these aforementioned regions, for spatial scales smaller than 0.5\,pc, the respective fragmentation lengths observed are consistent with the local spherical Jeans length. For spatial scales larger than 0.5\,pc, \citet{kainulainen13} find that the observed fragmentation scale is consistent with the prediction from the gravitational instability of a self-gravitating isothermal cylinder \citep[e.g.][]{nagasawa87}.  We explore next what physical processes could explain the smaller fragmentation length measured in OMC1\,1n from the separations of the individual cores.

Turbulence will produce a fragmentation scale larger than the Jeans length, as discussed in \citet{takahashi13}. Theoretical fragmentation length scales are directly proportional to the sound speed, $c_\mathrm{s}$. Turbulence can be included in this calculation by using an effective sound speed, $c_\mathrm{eff}$, instead of  $c_\mathrm{s}$; $c_\mathrm{eff}$ is given by $\sqrt{(\Delta v_\mathrm{t}^2+\Delta v_\mathrm{nt}^2)/8ln2}$, where $\Delta v_\mathrm{t}$ and $\Delta v_\mathrm{nt}$ correspond to the thermal and non-thermal velocity linewidths, respectively. Turbulence is included in the $\Delta v_\mathrm{nt}$ term. In the presence of turbulence, $c_\mathrm{eff} > c_\mathrm{s}$ and the fragmentation scale is consequently larger than the thermal Jeans length. Since the two measured fragmentation scales, $\lambda_\mathrm{obs}\_1$ and $\lambda_\mathrm{obs}\_2$, are either equal or smaller than their respective local Jeans lengths we can rule out turbulence pressure as a dominant physical process in the fragmentation of the filament and small-scale clumps, and conclude that it has essentially decayed for these size scales ($<$ 0.3\,pc) in OMC\,1n. This conclusion is consistent with that previously reached by \citet{houde09}, where they  find turbulence does not dominate the dynamics in OMC\,1. \citet{takahashi13} also rule out turbulence as a dominant physical process in the fragmentation of OMC\,3 and further suggest that it has dissipated for size scales $\leq$1\,pc.

We discuss next the role the magnetic field may play in the fragmentation of OMC\,1n.
Several polarimetric observations indicate that the magnetic field in OMC\,1n is mostly perpendicular to the filament and its small-scale clumps,  namely, far-infrared observations at  100\um\ from the Kuiper Airborne Observatory \citep{schleuning98}, submillimeter observations from CSO at 350\um\ \citep{schleuning98, houde04} and JCMT observations at 760\um\  \citep{vallee99}. The orientation of the magnetic field relative to the axis of the  filamentary cloud has important implications in terms of its stability \citep[e.g.][]{nagasawa87}. \citet{stodlokiewicz63} found that the critical length of perturbation increases with increasing magnetic field intensity, for the idealized case of a magnetic field perpendicular to an infinitely long cylinder of compressible gas. Assuming the orientation of the larger magnetic field is preserved for smaller scales in the filament \citep{li09}, and that the intensity of the magnetic field increases for smaller scales, it follows that the magnetic field topology of OMC\,1n cannot explain why $\lambda_{obs\_2}$ is smaller than $\lambda_{Jeans\_2}$ (Table \ref{tab:lengths}). Regarding the strength of the magnetic field, $B$,  \citet{houde04} calculate a total magnitude of $B\approx$850$\,\mu$G,  using the observed line-of-sight magnetic field strength of 360\,$\mu$G determined from Zeeman detection in CN \citet{crutcher99} and an inclination of 65\degr (consistent with a plane-of-the sky magnetic field strength of 760\,$\mu$G determined from polarimetric observations \citep{houde09}). They conclude that OMC\,1 is magnetically supercritical. The mass-to-flux ratio in units of the critical mass-to-flux ratio, $\gamma$, \citep{crutcher04} is given by:

\begin{equation}
\gamma=\frac{(M/\Phi)_{observed}}{(M/\Phi)_{critical}} = 7.6 \times 10^{-21} \frac{N(H_2)}{B}.
\label{eq:magnetic}
\end{equation}

\noindent Using the aforementioned value of $B$, and an $N(H_2)$ of $5 \times 10^{23}$ cm$^{-2}$ obtained from Herschel observations \citep{lombardi14}, we calculate $\gamma$ to be $\approx$ 4.5, and confirm that OMC\,1n is magnetically supercritical.  \citet{houde04} also calculate a magnetic-to-gravitational energy ratio of $\approx$ 0.26. Our current data do not allow us to analyse the magnetic field properties in more detail, however, from the above arguments it appears that the magentic field does not provide enough support against gravitational collapse, nor does it explain the shorter separations between cores observed in the small scale clumps.

Having ruled out turbulence and the magnetic field as responsible agents for the smaller fragmentation scales measured in OMC\,1n, $\lambda_{obs\_2}$, we now explore other possible processes. As already discussed, thermal fragmentation alone cannot fully explain $\lambda_{obs\_2}$, however, if thermal fragmentation were to occur concomitantly with the local collapse of the small-scale clumps, naturally the separations between the individual sources formed within the small-scale clumps are expected to decrease. \citet{Pon11} show that local collapse proceeds on a much smaller timescale than global collapse for filamentary structures. This analytical model would be consistent with the scenario where the local collapse in the small-scale clumps is occurring at a much faster pace than the global collapse of OMC\,1n, thus the shortening of the distances between sources would be more pronounced within the small-scale clumps, rather than in between them. We propose this may indeed be an important mechanism at play in OMC\,1n.  We plan to explore this scenario and constrain the effect that local vs. global collapse has on the final spatial separations of the sources with future ALMA data.

%
\section{Summary}
\label{sec:sum}

\begin{enumerate}

\item We found 24 new compact millimeter sources in OMC\,1n that are likely young protostars in the Class\,0 evolutionary phase, or pre-stellar objects.

\item The OMC\,1n sources range in size from 400\,au to 1300\,au, and range in mass from 0.1\msun\ to 2.3\msun\ (assuming a dust temperature of 15\,K and a distance to OMC\,1n of 414\,pc). These measured values are lower limits as our data is missing the zero spacing flux emitted by more extended structures.

\item The spatial distribution of the sources reveal a hierarchical two-level fragmentation the parental filamentary structure - the larger scale, $\lambda_\mathrm{obs\_1}$ = 31\arcsec $\pm$ 9\arcsec (0.06 $\pm$ 0.02\,pc), is associated with the fragmentation of the filament into small-scale clumps, while the smaller scale, $\lambda_\mathrm{obs\_2}$ = 6\arcsec $\pm$ 5\arcsec (0.01 $\pm$ 0.01\,pc), is associated with the fragmentation of the small-scale clumps into cores/SMA sources. 

\item Comparison of these two measured scales with the local Jeans lengths of the parental filament and of the small-scale clumps shows that the larger observed scale, $\lambda_\mathrm{obs\_1}$ is consistent with thermal fragmentation for inclinations greater than 50\degr. The smaller scale, $\lambda_\mathrm{obs\_2}$, is smaller than the local Jeans lengths of the parental small-scale clumps by a factor of $\sim$2-3.

\item We discuss other mechanisms that may explain the smaller measured scale, $\lambda_\mathrm{obs\_2}$. Further molecular line data is needed to disentangle the different physical processes involved, such as local collapse within OMC\,1n.

\end{enumerate}

\begin{acknowledgements}
The authors gratefully acknowledge an anonymous referee for their constructive review of an earlier version of this manuscript.
P.~S.~T. is very grateful for support from the Joint ALMA Observatory (Santiago, Chile) Science Visitor Programme while visiting co-author S. Takahashi. The authors wish to thank Jennifer Wiseman and Doug Johnstone for kindly sharing with us their VLA NH$_3$ and JCMT/SCUBA 850\um\ data, respectively, and for constructive comments. We would also like to thank Phil Myers, Josep Miquel Girart, Aina Palau, Andy Pon, \'Alvaro Hacar and Oliver Czoske for insightful discussions. We thank all the SMA staff members for making these observations possible. The authors wish to recognize and acknowledge the very significant cultural role and reverence that the summit of Mauna Kea has always had within the indigenous Hawaiian community.  We are most fortunate to have the opportunity to conduct observations from this mountain. P.~S.~T. also wishes to thank Alvin C. Inja for  positive interaction during the writing of this manuscript.
\end{acknowledgements}

\begin{appendix}
\section{Additional spatial correlation analysis}

We also perfomed a 2-point correlation analysis of the spatial distribution of the SMA sources, using the \citet{landy93} estimator . The correlation function, $\omega(\theta)$, is therefore given by:

\begin{equation}
\omega(\theta) = \frac{DD(\theta) - 2\cdot DR(\theta)+RR(\theta)}{RR(\theta)},
\label{eq:landy-szalay}
\end{equation}

\noindent where $DD(\theta)$ corresponds to the distribution of measured angular separations, $\theta$, between pairs of SMA sources. 

\begin{figure}[!h]
\centering
\includegraphics[width=\columnwidth]{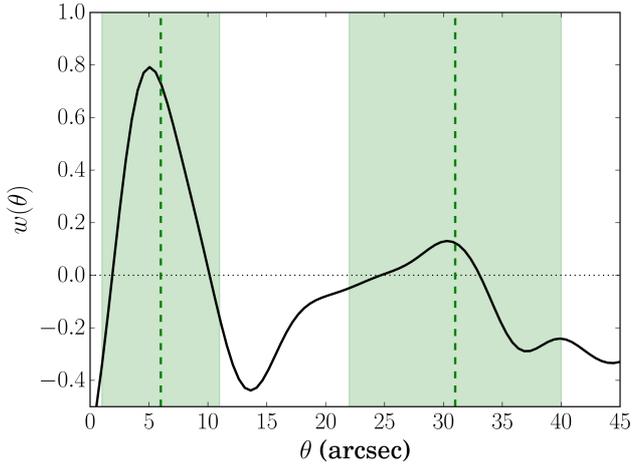}
\caption{Correlation function, $\omega(\theta)$, of the angular separation, $\theta$, between the SMA sources. The vertical dashed lines correspond to $\lambda_{obs\_1}$ and $\lambda_{obs\_2}$, respectively. The width of the shaded areas correspond to the corresponding error of $\lambda_{obs\_1}$ and $\lambda_{obs\_2}$. The dotted horizontal marks $\omega(\theta)$=0.}
\label{fig:correlation}
\end{figure}

\noindent $RR(\theta)$ corresponds to the distribution of angular separations between pairs of random sources, and $DR(\theta)$ corresponds to the distribution of angular separations between pairs of an SMA source and a random source.
The sample of random sources consisted of the same number of sources as the sample of SMA sources, placed within the same RA and DEC ranges, and was generated 10\,000 times; the final $RR(\theta)$ and $DR(\theta)$ distributions were averaged along $\theta$. \\

Figure \ref{fig:correlation} shows that $\omega(\theta)$ has two maxima, denoting preferred spatial correlations at separations consistent with $\lambda_{obs\_1}$ and $\lambda_{obs\_2}$, as measured above using the median 1st and 7th nearest neighbor separations of the SMA sources, respectively.

\end{appendix}

\end{document}